\documentclass[prd,showpacs,preprintnumbers,amsmath,amssymb]{revtex4}

\usepackage{graphicx}
\usepackage{dcolumn}
\usepackage{bm}

\begin{document}

\title{Dark Energy Regulation with Approximate Emergent Conformal Symmetry}

\author{Yongsung Yoon}\email{cem@hanyang.ac.kr}
\affiliation{Physics Department, Research Institute for Natural Sciences, Hanyang University, Seoul 133-791, Korea}

\date{\today}

\begin{abstract}
A cosmic potential which can relax the vacuum energy is proposed in a framework of scalar-tensor gravity. In the phase
of the gravity scalar field around the evolution with an approximate emergent conformal symmetry, we have obtained a
set of cosmological equations with the dark energy regulated to the order of a conformal anomaly parameter. Through a
role of the cosmic potential, the vacuum energy which could be generated in matter Lagrangian does not contribute to
the dark energy in the phase.

\end{abstract}
\pacs{95.36.+x, 95.30.Sf, 04.20.-q, 98.80.Jk} \maketitle

\section{Introduction}

Since the development of General Relativity and modern cosmology, the cosmological constant problem, why the
cosmological constant is almost zero, has been a big mystery, because there has been no successful mechanism to remove
the vacuum energy which could be generated by quantum fields. The arguments with super-symmetry(or scale symmetry) can
prevent the emergence of a non-vanishing cosmological constant only when the symmetry is unbroken. However, the energy
scale of such a symmetry breaking seems to be quite large so that still a large vacuum energy could be generated below
the energy scale.

Making the problem even harder, recent astronomical observations of a late time acceleration of our universe
\cite{ACC1,ACC2} have added another puzzle to the cosmological constant problem: the observational dark energy density
is not zero, but as small as $\rho^{obs}_{\Lambda} \sim (10^{-3}{\rm eV})^{4}$, which is a comparable order of
magnitude to the matter density of our present universe. With any conceivable energy cut-off and symmetry breaking
scale, the vacuum energy of quantum fields would contribute much larger to the cosmological constant than the
observations (For a recent review, see Refs. \cite{I1,I2} and references therein).

Thus, on top of the old puzzle for dark energy, how the vacuum energy in matter Lagrangian does not contribute to the
dark energy, we are now facing another new puzzle, why the dark energy density is so close to the matter density at
present.

As yet, there has been no successful mechanism to remove the vacuum energy leaving such a small dark energy density
which is close to the matter density in the present universe. Since there is no room to conceive a mechanism to remove
or relax the vacuum energy in General Relativity, we consider a scalar-tensor gravity model which has an additional
gravity scalar field in gravity sector like Brans-Dicke gravity \cite{BD1,BD2} or the induced gravity
\cite{IGO1,IGO2,IGO3,IGO4,IGO5,Odint}. The low energy limit of string theory \cite{D-brane,string} and Kaluza-Klein
theory \cite{KK5} with a proper spacetime compactification also give a scalar-tensor gravity action with specific
scalar couplings. In scalar-tensor gravity models, the gravity scalar is a part of gravity fields and its dynamics is
quite different from those of ordinary scalar fields which are driven by their potential gradients.

In any scalar-tensor gravity models which can give a flat spacetime solution with a constant scalar field in vacuum,
there is no hope to find an exact cancellation mechanism of the cosmological constant as argued by Weinberg
\cite{No-go}. However, with an evolving gravity scalar field, we have found a regulation mechanism of dark energy in a
scalar-tensor gravity model with a unique cosmic scalar potential. Even though the scalar-tensor gravity model does not
give a flat spacetime for a constant gravity scalar field even in vacuum, it is found that the non-flatness could be
quite negligible in a non-cosmic scale.

\section{Scalar-Tensor Gravity}

We consider the following scalar-tensor gravity action introducing a cosmic scalar potential $V(\Phi)$ which is to be
determined by a relax mechanism to work,
\begin{eqnarray}
S= -\int d^{4}x\sqrt{g}[\frac{1}{2}\Phi^{2} R +\frac{\xi}{2}(\partial \Phi)^{2} + V(\Phi)] + \int d^{4}x\sqrt{g} {\cal
L}_{m}, \label{action}
\end{eqnarray}
where ${\cal L}_{m}$ is the matter Lagrangian, and $\xi$ is a dimensionless coupling constant between the gravity
scalar field and the metric. $\xi=6$ is for the conformal coupling \cite{YP2}, and $\xi=4$ is for the low energy limit
of string theory (assuming a Ricci-flat compactification of the internal $(D-4)$ dimensional space decoupled from our
4-dimensional spacetime) with the field identification
\begin{eqnarray}
\Phi \equiv \frac{1}{\kappa_{0}} e^{-\phi} \label{ReDef}
\end{eqnarray}
in the string frame \cite{D-brane,string}.

In vacuum without ${\cal L}_{m}$, it would be convenient to choose the Einstein frame redefining the metric,
$g_{\mu\nu} \rightarrow \Phi^{-2}g_{\mu\nu}$. However, in presence of matter, the frame change of metric makes
couplings of the gravity scalar field to matter fields different depending on their spins in the matter Lagrangian,
which gives a complicate matter energy-momentum tensor depending on the gravity scalar field in the new frame.

The equations of motion for the gravity scalar field and the metric can be written as
\begin{eqnarray}
\xi\nabla^{2}\Phi = \Phi R + \frac{\partial V(\Phi)}{\partial \Phi} , \label{S-eq}
\end{eqnarray}
\begin{eqnarray}
\Phi^{2} G_{\mu\nu} - 2\Phi\nabla_{\mu}\nabla_{\nu}\Phi +2g_{\mu\nu}\Phi\nabla^{2}\Phi
+(\xi-2)\nabla_{\mu}\Phi \nabla_{\nu}\Phi - \frac{\xi-4}{2}g_{\mu\nu}(\nabla\Phi)^{2} - g_{\mu\nu}V(\Phi) =
T^{(m)}_{\mu\nu} , \label{G-eq}
\end{eqnarray}
where $T^{(m)}_{\mu\nu}$ is the matter energy-momentum tensor. Combining the trace part of equation (\ref{G-eq}) and
the gravity scalar equation (\ref{S-eq}), we can rewrite the gravity scalar equation equivalently as
\begin{eqnarray}
(\xi-6)[\Phi \nabla^{2}\Phi + (\nabla\Phi)^{2}] +T^{(m)} +4V(\Phi) -\Phi\frac{\partial V(\Phi)}{\partial \Phi} = 0 ,
\label{GS-eq}
\end{eqnarray}
where $T^{(m)}$ is the trace of the matter energy-momentum tensor.

As we can see from Eq.(\ref{GS-eq}), the driving force of the gravity scalar field is given by the conformal anomaly
$T^{(m)} +4V -\Phi V'$ of the whole system, which is quite distinctive from an ordinary scalar field whose driving
force is the gradient of its potential. Thus, the gravity scalar field can not be settled down at a potential minimum
$V(\Phi)'=0$ unlike an ordinary scalar field, but be dynamically driven to the emergent conformal symmetry with
$T^{(m)} +4V -\Phi V'=0$.

Due to the presence of the cosmic scalar potential $V(\Phi)$, the equations Eqs.(\ref{S-eq},\ref{G-eq}) do not allow
the exact flat spacetime solution for a constant gravity scalar field even in vacuum. However, in a non-cosmic scale,
it is found that the deviation from the flat spacetime could be negligible provided that the cosmic potential satisfies
a certain condition.

\section{Cosmological Evolution}

To examine a cosmological evolution, we consider a homogeneous gravity scalar field $\Phi(t)$ with only time
dependence, and adopt the Robertson-Walker metric with a vanishing spatial curvature($k=0$),
\begin{eqnarray}
ds^{2}=dt^{2}-S^{2}(t)[dr^{2}+r^{2}d\Omega^{2}]. \label{metric}
\end{eqnarray}

Then, Eqs.(\ref{S-eq},\ref{G-eq}) are reduced to a set of cosmological equations as shown below. (For various
cosmological solutions in scalar-tensor gravity models with specific scalar potentials, see
Ref.\cite{IG0,IG1,IG2,IG3,IG4,IG5,IG6,IG7,IG8,IG9,IGa,Odin}) With the over-dot denoting a time derivative,

\begin{eqnarray}
3\Phi^{2}H^{2} +6H\Phi\dot{\Phi} +\frac{\xi}{2}\dot{\Phi}^{2} = \rho_{m} + \rho_{\Lambda}(\Phi) , \label{CosG00-eq}
\end{eqnarray}
\begin{eqnarray}
2 a \Phi^{2} +H^{2}\Phi^{2} +4H\Phi\dot{\Phi} +2\Phi\ddot{\Phi} - \frac{\xi-4}{2}\dot{\Phi}^{2} = -p_{m} +
\rho_{\Lambda}(\Phi) , \label{CosG11-eq}
\end{eqnarray}
\begin{eqnarray}
\xi(\ddot{\Phi}+3H\dot{\Phi}) + 6\Phi(H^{2}+a) = \frac{\partial V(\Phi)}{\partial \Phi} , \label{CosD-eq}
\end{eqnarray}
where $H \equiv \dot{S}/S$, $~a \equiv \ddot{S}/S$, $~T^{(m)0}_{0}=\rho_{m}+\rho_{vac}$,
$~T^{(m)1}_{1}=T^{(m)2}_{2}=T^{(m)3}_{3}=-p_{m}+\rho_{vac}$.  $\rho_{vac}$ is a vacuum energy which could be generated
by quantum effects in the matter Lagrangian. We can see that the sum of the vacuum energy $\rho_{vac}$ in the matter
Lagrangian and the cosmic potential $V(\Phi)$ plays the role of the total dark energy density, $\rho_{\Lambda}(\Phi)
\equiv \rho_{vac} + V(\Phi)$.

Obviously, Eqs.(\ref{CosG00-eq},\ref{CosG11-eq},\ref{CosD-eq}) do not admit an exact constant solution of the gravity
scalar field in the presence of matter. In a slowly varying phase of the gravity scalar field such that
\begin{eqnarray}
\mid \frac{\dot{\Phi_{s}}}{\Phi_{s} H} \mid \ll 1 , \quad \mid \frac{\ddot{\Phi_{s}}}{\Phi_{s} H^{2}} \mid \ll 1 ,
\label{slow}
\end{eqnarray}
which would be a late time phase of the cosmological evolution with a very small matter density $\rho_{m}$, the above
cosmological equations (\ref{CosG00-eq},\ref{CosG11-eq},\ref{CosD-eq}) can be approximated to the well-known Friedmann
equations (\ref{F1-eq},\ref{F2-eq}) and an additional scalar equation (\ref{C0-eq}).
\begin{eqnarray}
H^{2} \simeq \frac{8\pi G(\Phi_{s})}{3}\rho_{T} , \label{F1-eq}
\end{eqnarray}
\begin{eqnarray}
a \simeq -\frac{4\pi G(\Phi_{s})}{3}(\rho_{T} +3p_{T}) , \label{F2-eq}
\end{eqnarray}
\begin{eqnarray}
\rho_{m} - 3p_{m} +4\rho_{vac} +4V(\Phi_{s}) \simeq \Phi_{s}\frac{\partial V(\Phi)}{\partial \Phi}\mid_{\Phi_{s}} ,
\label{C0-eq}
\end{eqnarray}
where total energy density is $\rho_{T} \equiv \rho_{m} + \rho_{\Lambda}(\Phi_{s})$, total pressure is $~p_{T} \equiv
p_{m} -\rho_{\Lambda}(\Phi_{s})$, and the effective Newton's constant is $G(\Phi_{s}) \equiv 1/(8\pi\Phi_{s}^{2})$. It
is found that, in the slow phase of the gravity scalar field, the scalar equation (\ref{CosD-eq}) becomes a trace
anomaly relation (\ref{C0-eq}), whose left hand side is the total trace anomaly of the universe.

\section{A Cosmic Potential to Relax the Vacuum Energy in the Slow Phase}

We try to find a cosmic potential which can relax the vacuum energy in matter Lagrangian in the slow phase of the
gravity scalar field. For a given vacuum energy $\rho_{vac}$ in matter Lagrangian, the scalar equation (\ref{CosD-eq})
in the phase is satisfied for $\Phi_{s}$ as Eq.(\ref{C0-eq}). For a different vacuum energy $\rho'_{vac}$, the equation
(\ref{C0-eq}) would be satisfied with a different gravity scalar field $\Phi'_{s}$ instead of $\Phi_{s}$ as
\begin{eqnarray}
\rho_{m} - 3p_{m} +4\rho'_{vac} +4V(\Phi'_{s})  \simeq \Phi'_{s}\frac{\partial V(\Phi)}{\partial \Phi}\mid_{\Phi'_{s}}
. \label{C1-eq}
\end{eqnarray}
For these two different vacuum energies $\rho_{vac}$ and $\rho'_{vac}$, the difference in the total dark energy density
 in the slow phase is found as
\begin{eqnarray}
\Delta\rho_{\Lambda} =\rho_{\Lambda}(\Phi'_{s}) -\rho_{\Lambda}(\Phi_{s}) =
V(\Phi'_{s})-V(\Phi_{s})+\rho'_{vac}-\rho_{vac} \simeq \frac{1}{4}[\Phi'_{s}\frac{\partial V(\Phi)}{\partial
\Phi}\mid_{\Phi'_{s}} - \Phi_{s}\frac{\partial V(\Phi)}{\partial \Phi}\mid_{\Phi_{s}}] . \label{Zero-eq}
\end{eqnarray}

Thus, it is found that we can have the same order of total dark energy density $\rho_{\Lambda}$ in the slow phase,
whatever the vacuum energy $\rho_{vac}$ is in matter Lagrangian provided that the cosmic scalar potential satisfies the
equation, $\Phi\partial V(\Phi)/\partial\Phi = \alpha$ for a constant $\alpha$. Thus we choose the cosmic scalar
potential uniquely as;
\begin{eqnarray}
V(\Phi) = \frac{\alpha}{2}\ln(\frac{\Phi}{\mu})^{2} . \label{Pot-eq}
\end{eqnarray}
Here $\alpha$ is a dimensionful constant which would have some other physical origin. However, both the dimensionful
constant $\mu$ and the vacuum energy $\rho_{vac}$ in matter Lagrangian hide into the total dark energy density and do
not appear as independent physical quantities;
\begin{eqnarray}
\rho_{\Lambda}(\Phi) = \rho_{vac} + \frac{\alpha}{2}\ln(\frac{\Phi}{\mu})^{2} . \label{DE-eq}
\end{eqnarray}

For the cosmic potential Eq.(\ref{Pot-eq}), the total dark energy density $\rho_{\Lambda}$ satisfies the following
trace anomaly relation with a matter density $\rho_{m}$ in the slow phase of the gravity scalar field $\Phi_{s}$,
\begin{eqnarray}
\rho_{m} - 3p_{m} + 4\rho_{\Lambda}(\Phi_{s}) \simeq \alpha . \label{CR-eq}
\end{eqnarray}
The left hand side of this equation is the trace of the total energy momentum tensor including all matter and total
dark energy of our universe. Thus, Eq.(\ref{CR-eq}) is a trace anomaly relation, which states that the total trace
anomaly of our universe becomes approximately a constant $\alpha$ in the slow phase of the gravity scalar field.
Because the left hand side of (\ref{CR-eq}) is the total trace anomaly, we might speculate that the right hand side
$\alpha$ would also have an origin of a conformal anomaly. The possibility seems to be plausible observing that the
logarithmic cosmic potential (\ref{Pot-eq}) is the linear form of coupling $\phi\Theta^{\mu}_{\mu}$ between a scalar
field $\phi$ and a conformal anomaly $\Theta^{\mu}_{\mu}=\alpha$ which might have an origin in quantum chromodynamics
considered in \cite{No-go,anomal} with the scalar field redefinition (\ref{ReDef}). Thus, we may call $\alpha$ as a
conformal anomaly parameter.

Plugging the cosmic potential (\ref{Pot-eq}) into the full equations
Eqs.(\ref{CosG00-eq},\ref{CosG11-eq},\ref{CosD-eq}) and combining them, we have the modified energy conservation
equation
\begin{eqnarray}
\frac{d}{dt}(\rho_{T}\frac{4}{3}\pi S^{3}) + p_{T}\frac{d}{dt}(\frac{4}{3}\pi S^{3}) =
\frac{dV(\Phi)}{dt}(\frac{4}{3}\pi S^{3}) = \alpha\frac{\dot{\Phi}}{\Phi}(\frac{4}{3}\pi S^{3}), \label{energy-eq}
\end{eqnarray}
and the exact form of the trace anomaly relation (\ref{CR-eq})
\begin{eqnarray}
\rho_{m} -3p_{m}  +4\rho_{\Lambda}(\Phi) + (\xi-6)(\Phi\ddot{\Phi} +3H\Phi\dot{\Phi} +\dot{\Phi}^{2}) =
\Phi\frac{\partial V(\Phi)}{\partial \Phi} = \alpha .  \label{Comb-eq}
\end{eqnarray}
From Eq.(\ref{energy-eq}), using the total energy density $\rho_{T} \equiv \rho_{m} + \rho_{\Lambda}(\Phi)$, the total
pressure $~p_{T} \equiv p_{m} -\rho_{\Lambda}(\Phi)$ with the explicit form Eq.(\ref{DE-eq}) of $\rho_{\Lambda}(\Phi)$,
it is found that the matter energy conservation holds exactly regardless of the gravity scalar field $\Phi$,
\begin{eqnarray}
\frac{d}{dt}\rho_{m} + 3H(\rho_{m}+p_{m}) = 0 . \label{matter-eq}
\end{eqnarray}

Especially, in vacuum without any matter such that $\rho_{m}=0=p_{m}$, which would be the case of our universe in the
infinite future, we have the Hubble parameter, and the acceleration parameter in vacuum for a constant gravity scalar
field $\Phi_{vac}$ as
\begin{eqnarray}
H^{2}_{vac} = \frac{\alpha}{12\Phi^{2}_{vac}}=a, ~\rho_{\Lambda}(\Phi_{vac})=\frac{\alpha}{4} \label{vacuum} .
\end{eqnarray}
Thus, the ratio $(a/H^{2})_{vac} =1$ in vacuum is a little larger than the current observational value $(a/H^{2})_{obs}
\approx 0.55$ \cite{ACC1,ACC2,LCDM}.

In the late time evolution of our universe dominated by cold matter with the matter pressure $p_{m}=0$, the matter
energy conservation equation (\ref{matter-eq}) tells $\rho_{m} = A/S^{3}(t)$ with a constant $A$. Let us assume that
the trace anomaly relation (\ref{CR-eq}) holds for $\Phi_{s}$ as
\begin{eqnarray}
\rho_{m} - 3p_{m} + 4\rho_{\Lambda}(\Phi_{s})= \alpha \label{exact} ,
\end{eqnarray}
which is the evolution with the emergent conformal symmetry.
Then, taking a time derivative of equation (\ref{exact})
with the help of the explicit form Eq.(\ref{DE-eq}), we find that
\begin{eqnarray}
\frac{\dot{\Phi}_{s}}{H\Phi_{s}} = \frac{3}{4}\frac{\rho_{m}}{\alpha}, \quad \frac{\ddot{\Phi}_{s}}{H^{2}\Phi_{s}}=
\frac{3}{4}(\frac{a}{H^{2}}-4)\frac{\rho_{m}}{\alpha}+\frac{9}{16}(\frac{\rho_{m}}{\alpha})^{2} \label{stable-sol} .
\end{eqnarray}

Therefore, the slowly varying assumption (\ref{slow}) is only possible when the matter density ratio $\rho_{m}/\alpha
\ll 1$ is very small.

Let us examine the stability of $\Phi(t)$ around the slowly varying $\Phi_{s}(t)$. With
$\Phi(t)=\Phi_{s}(t)+\varphi(t)$, Eq.(\ref{Comb-eq}) can be written, up to linear terms of $\varphi(t)$, as
\begin{eqnarray}
\ddot{\varphi} = -(3H+2\frac{\dot{\Phi}_{s}}{\Phi_{s}})\dot{\varphi} +
\frac{1}{\Phi_{s}^{2}}(\dot{\Phi}_{s}^{2}-\frac{4\alpha}{\xi-6})\varphi
 \label{stability} ,
\end{eqnarray}
where the first term in the right hand side of Eq.(\ref{stability}) is a frictional term. From Eq.(\ref{stable-sol}),
we find that
\begin{eqnarray}
\dot{\Phi}_{s}^{2} = \frac{9}{16}(\frac{\rho_{m}}{\alpha})^{2}H^{2}\Phi_{s}^{2} \simeq
(\frac{\rho_{m}}{\alpha})^{2}\alpha,
\end{eqnarray}
Thus, if $\xi < 6$ or the matter density ratio $\rho_{m}/\alpha$ is large so that $\dot{\Phi}_{s}^{2} >
\frac{4\alpha}{\xi-6}$ as in the early stage of the cosmological evolution, the evolution $\Phi(t)$ around
$\Phi_{s}(t)$ is not stable. However, if $\xi > 6$, the evolution $\Phi(t)$ around $\Phi_{s}(t)$ is stable in the late
time evolution with a small matter density ratio so that $\dot{\Phi}_{s}^{2} < 4\alpha/(\xi-6)$.

Thus, we have found that, once $\Phi(t)$ satisfies the trace anomaly relation (\ref{exact}), it stays around
$\Phi_{s}(t)$ of having the emergent conformal symmetry for $\xi \geq 6$. With the help of the slowly varying
$\Phi_{s}$ (\ref{stable-sol}), from Eq.(\ref{Comb-eq}), we obtain an approximate relation for the evolution $\Phi(t)$
around $\Phi_{s}(t)$ as
\begin{eqnarray}
\rho_{m} - 3p_{m} + 4\rho_{\Lambda}(\Phi)- \alpha \cong (\xi-6)\left( 3-\frac{9}{4}-\frac{9}{8}\frac{\rho_{m}}{\alpha}
-\frac{3}{4}\frac{a}{H^{2}} \right) (\frac{\rho_{m}}{\alpha})H^{2}\Phi^{2} . \label{deviation}
\end{eqnarray}
Therefore, through the evolution $\Phi(t)$ around $\Phi_{s}(t)$, the dark energy density is regulated to the order of
the conformal anomaly parameter $\alpha$ because the right hand side of Eq.(\ref{deviation}) is an order of
$(\frac{\rho_{m}}{\alpha})H^{2}\Phi_{s}^{2}$ with a relatively small matter density ratio $\frac{\rho_{m}}{\alpha}$ in
the late time cosmological evolution. In the phase of the evolution around $\Phi_{s}(t)$ with the emergent conformal
symmetry, the cosmological equations (\ref{CosG00-eq},\ref{CosG11-eq}) can be approximated up to the linear terms of
$\rho_{m}/\alpha$ as
\begin{eqnarray}
3\Phi_{s}^{2}H^{2} \simeq \frac{\rho_{m}+\rho_{\Lambda}}{1+\frac{3\rho_{m}}{2\alpha}} \label{Cos-A1-eq} ,
\end{eqnarray}
\begin{eqnarray}
6\Phi_{s}^{2}a \simeq \frac{3(\rho_{\Lambda}-p_{m})}{1+\frac{3\rho_{m}}{4\alpha}} - \frac{(\rho_{m}+\rho_{\Lambda})
(1-\frac{3\rho_{m}}{\alpha})}{(1+\frac{3\rho_{m}}{4\alpha})(1+\frac{3\rho_{m}}{2\alpha})} \label{Cos-A2-eq} ,
\end{eqnarray}
which are independent of $\xi$ provided that $\xi$ is not much larger than $6$.

\section{Conclusions and Discussion}

In the late time cosmological evolution, through a slow evolution of the gravity scalar field satisfying the trace
anomaly relation (\ref{CR-eq}) approximately, the dark energy density $\rho_{\Lambda}$ is regulated to the order of the
conformal anomaly parameter $\alpha$. Because the left hand side of (\ref{CR-eq}) is the total trace anomaly, we might
speculate that the right hand side $\alpha$ would also have an origin of a conformal anomaly. The possibility seems to
be plausible observing that the cosmic potential (\ref{Pot-eq}) is the linear form of coupling $\phi\Theta^{\mu}_{\mu}$
between a scalar field $\phi$ and a conformal anomaly $\Theta^{\mu}_{\mu}=\alpha$ which might have an origin in quantum
chromodynamics considered in \cite{No-go,anomal} with the scalar field redefinition (\ref{ReDef}). Assuming that our
universe is currently near the slow phase of the gravity scalar field $\Phi_{s}(t)$, the trace anomaly relation
(\ref{CR-eq}) implies that the conformal anomaly parameter $\alpha$ should be the order of the dark energy density
$\rho_{\Lambda}$ in our present universe. The conformal anomaly parameter $\alpha$ would give a new fundamental length
scale of gravity instead of Newton constant as $\alpha \sim \rho_{\Lambda} \sim (10^{-3}eV)^{4} \sim (0.1mm)^{4}$. Due
to the smallness of the conformal anomaly parameter $\alpha$, the approximate flatness of the vacuum spacetime in a
non-cosmic scale is secure, and the local gravity in a solar system scale is not affected by the presence of the
logarithmic cosmic potential (\ref{Pot-eq}).

Recently, the ratio of the cosmic acceleration parameter $a$ to the Hubble parameter $H^{2}$ was determined through the
help of Type-I supernovae as standard candles \cite{ACC1,ACC2,LCDM},
\begin{eqnarray}
\left(\frac{a}{H^{2}}\right)_{exp} \simeq 0.55 \label{ratio-exp}.
\end{eqnarray}
From the ratio of Eq.(\ref{Cos-A2-eq}) to Eq.(\ref{Cos-A1-eq}), $\frac{a}{H^{2}}$ with $p_{m}=0$ (it is believed that
the matter in our current universe is cold \cite{LCDM}), we can find a solution of $\rho_{m}/\alpha$ which satisfies
the experimental ratio $(a/H^{2})_{exp}$ numerically with a given $\rho_{\Lambda}$. If we use the experimental value
(\ref{ratio-exp}) with $\rho_{\Lambda}=0.1\alpha$ as an example, then we find the corresponding solution
$\rho_{m}/\alpha=0.07$. For this $\rho_{m}/\alpha$, the cosmological equations (\ref{Cos-A1-eq},\ref{Cos-A2-eq})
deviates a little from the Friedmann equations, giving the matter density $\rho_{m}$, the dark energy density
$\rho_{\Lambda}$, the critical density $\rho_{c}$ at present as
\begin{eqnarray}
\rho_{m}=0.07\alpha, \quad \rho_{\Lambda}=0.1\alpha, \quad \rho_{c} \equiv 3\Phi^{2}H^{2} = \frac{3}{8\pi G}H^{2}
=0.15\alpha  \label{fit-exp1} .
\end{eqnarray}
The trace anomaly relation holds approximately as $\rho_{m}+4\rho_{\Lambda} \simeq 0.29\alpha$ for this
$\rho_{m}/\alpha$. From the difference between the infinite future and the current total dark energy densities, we can
estimate the increase of gravity scalar field $\Phi_{\infty}^{2} \simeq 1.3\Phi^{2}_{now}$ and the corresponding
decrease of gravitational constant $G_{N\infty} \simeq 0.74 G_{N,now}$ in the infinite future.

Even though the effect of the cosmic potential in a galactic scale might need further exploration, the relax mechanism
of the vacuum energy by a gravity scalar field with the cosmic potential proposed may be a candidate to resolve the
puzzles for the dark energy.




\begin{thebibliography}{99}


\bibitem{ACC1} A.G. Riess {\it et al.}, Astron. J. {\bf 116}, 1009 (1998); Astron. J. {\bf 118}, 2668
(1999); Astron. J. {\bf 607}, 665 (2004); S. Perlmutter {\it et al.}, Nature {\bf 398}, 51 (1998); Astron.
J. {\bf 517}, 565 (1999); R. Knop {\it et al.}, Astron. J. {\bf 598}, 102 (2003); B. Barris {\it et al.},
Astron. J. {\bf 602}, 571 (2004);

\bibitem{ACC2} S. Perlmutter {\it et al.}, Astrophys. J. {\bf 517}, 565 (1999).

\bibitem{I1} A. Silvestri and M. Trodden, Rept. Prog. Theor. {\bf 72}, 096901 (2009).

\bibitem{I2} M. Sami, arXiv:0904.3445, (2009).

\bibitem{BD1} C.H. Brans and R.H. Dicke, Phys. Rev. {\bf 124}, 925 (1961).

\bibitem{BD2} P.G. Bergmann, Int. J. Theor. Phys. {\bf 1}, 25 (1968).

\bibitem{IGO1} A. Zee, Phys. Rev. Lett. {\bf 42}, 417 (1979).

\bibitem{IGO2} L. Smolin, Nucl. Phys. {\bf B160}, 253 (1979).

\bibitem{IGO3} A. Adler, Rev. Mod. Phys. {\bf 54}, 729 (1982).

\bibitem{IGO4} A.D. Sakharov. Dokl. Akad. Nauk. SSSR 117, 70 (1967) [Sov. Phys. Dokl. {\bf 12}, 1040
(1967)].

\bibitem{Odint} I.L. Buchbinder and S.D. Odintsov, Int. J. Mod. Phys. A {\bf 3}, 1859 (1988); Yad.Fiz. {\bf 46}, 1233
(1987).

\bibitem{IGO5} P. Batra, K. Hinterbichler, L. Hui, and D. Kabat, Phys. Rev. {\bf D78}, 043507 (2008).

\bibitem{D-brane} C.V. Johnson, {\it D-brane primer}, Boulder 1999, Strings, branes and gravity, 129-350
[arXiv:hep-th/0007170].

\bibitem{string} J. Polchinski, {\it String Theory}, Vols. I and II, Cambridge University Press (1998).

\bibitem{KK5} F. Darabi, Mod. Phys. Lett. {\bf A25}, 1635 (2010); K.Bamba, S. Nojiri, and S.D. Odintsov, arXiv:1304.6191.

\bibitem{No-go} S. Weinberg, Rev. Mod. Phys. {\bf 61}, 1 (1989).

\bibitem{YP2} Y. Yoon, Phys. Rev. {\bf D59}, 127501 (1999); C.J. Park and Y. Yoon, Gen. Rel. Grav. {\bf 29}, 765
(1997).

\bibitem{IG0} T. Padmanabhan, {\it Quantum Conformal Fluctuations, Induced Gravity And Cosmology}, CERN
Preprint:CERN-TH-3706 (Sep. 1983).

\bibitem{IG1} F.S. Accetta, D.J. Zoller, and M.S. Turner, Phys. Rev. {\bf D31}, 3046 (1985).

\bibitem{IG2} F.S. Accetta and J.J. Trester, Phys. Rev. {\bf D39}, 2854 (1989).

\bibitem{IG3} E. Carugno, S. Capozziello, and F. Occhionero, Phys. Rev. {\bf D47}, 4261 (1993).

\bibitem{IG4} D. La, Phys. Rev. {\bf D44}, 1680 (1991); {\it The "Scaled" Induced-Gravity Inflationary
Cosmology}, U.C. Berkely Preprint:CfPA-TH-90-015A (Jul. 1990).

\bibitem{IG5} J.L. Cervantes-Cota and H. Dehnen, Nucl. Phys. B {\bf 442}, 391 (1995).

\bibitem{IG6} W.F. Kao, Phys. Lett. A {\bf 147}, 165 (1990).

\bibitem{IG7} R. Fakir and  W.G. Unruh, Phys. Rev. D {\bf 41}, 1792 (1990).

\bibitem{IG8} D.I. Kaiser, Phys. Rev. {\bf D49}, 6347 (1994).

\bibitem{IG9} V. Faraoni, Class. Quant. Grav. {\bf 26}, 145014 (2009).

\bibitem{IGa} A. Cerioni, F. Finelli, A. Tronconi, and G. Venturi, Phys. Lett. B {\bf 681}, 383 (2009).

\bibitem{Odin} E. Elizalde, S. Nojiri, and S.D. Odintsov, Phys. Rev. D {\bf 70},  043539(2004).

\bibitem{anomal} R.D. Peccei, J. Sol$\grave{a}$, and C. Wetterich, Phys. Lett. B {\bf 195}, 183 (1987).

\bibitem{LCDM} S.M. Carroll, Liv. Rev. Rel. {\bf 4}, 1 (2001).



\end{thebibliography}
\end{document}